\documentclass[11pt]{article}
\usepackage{BSMvietnam,epsfig}

\def\be{\begin{equation}}
\def\ee{\end{equation}}
\def\bea{\begin{eqnarray}}
\def\eea{\end{eqnarray}}

\begin{document}
\vspace*{4cm}
\title{UNIFYING ASYMMETRIC INERT FERMION DOUBLET DARK MATTER AND LEPTOGENESIS WITH NEUTRINO MASS}

\author{ NARENDRA SAHU }

\address{Department of Physics, Indian Institute of Technology Hyderabad\\
Yeddumailaram 502205, AP, India}

\maketitle

\abstracts{We propose a scalar Triplet extension of the standard model (SM) to unify 
the origin of neutrino mass with the visible and dark matter component of the Universe. We 
assume that the scalar triplet is super heavy, so that its CP-violating out-of-equilibrium 
decay in the early Universe not only produce asymmetric dark matter which is the neutral 
component of an additional vector like fermion doublet, but also give rise to lepton asymmetry. 
The latter gets converted to observed baryon asymmetry via B+L violating sphaleron processes. Below electroweak 
phase transition the scalar triplet acquires a vacuum expectation value and give rise to sub-eV Majorana 
masses to three flavors of active neutrinos. Thus an unification of the origin of neutrino mass, lepton asymmetry 
and asymmetric dark matter is achieved within a scalar triplet extension of the SM.}

\section{Introduction}
Strong evidence from galaxy rotation curve, gravitational lensing and large scale structure implies that there 
exist invisible matter whose relic abundance,  $\Omega_{\rm DM} \sim 0.23$, is well measured by the WMAP 
satellite~\cite{Komatsu:2010fb}. This invisible matter is usually called dark matter (DM) and interacts gravitationally 
as the above evidences imply. However, it does not have any electromagnetic interaction with the visible matter. But DM 
can have weak interaction as many direct, indirect and collider experiments are currently exploring it. 

The existing evidence of DM imply that it should be massive, electrically neutral and is stable on 
cosmological time scales. However, such a particle is absent in the SM particle spectrum and 
therefore led to many theories in the physics beyond SM scenarios. Another issue concerning SM is the origin of tiny amount 
of visible matter in the Universe which is in the form of baryons with $\Omega_{\rm B} \sim 0.04$, that could be arising from 
a baryon asymmetry $n_B/n_\gamma \sim 6.15 \times 10^{-10}$, as established by WMAP combined with the big-bang nucleosynthesis 
(BBN) measurements. Moreover, the tiny masses of three active neutrinos required by the oscillation experiments are not explained 
within the SM framework. 

The fact that $\Omega_{\rm DM}\sim 5 \Omega_{\rm B}$ could be a hint that both sectors share a common origin and the present 
relic density of DM is also generated by an asymmetry~\cite{Kohri:2009yn,Arina:2011cu,Arina:2012fb,Arina:2012aj}. In this talk 
we present an economic model where the asymmetric DM and baryonic matter densities can be generated simultaneously in the early 
Universe (above electroweak phase transition) and Majorana masses of three flavours of active neutrinos at late epochs 
(below electroweak phase transition) from a common source. We extend the SM by introducing two scalar triplets 
$\Delta_1$ and $\Delta_2$. We also add a vector like fermion doublet $\psi$ and impose a discrete 
symmetry $Z_2$ under which $\psi$ is odd while rest of the fields are even. As a result the neutral component of 
$\psi$ becomes a candidate of DM. The CP-violating out-of-equilibrium decays: $\Delta_i \to LL$, where $L$ is the $SU(2)_L$ 
lepton doublet, and $\Delta_i \to \psi\psi$ then induce the asymmetries simultaneously in visible and DM 
sectors. The lepton asymmetry is then transferred to a baryon asymmetry through the B+L violating sphaleron 
transitions while the asymmetry in the DM sector remains intact as the B+L current of vector like fermion 
doublet $\psi$ is anomaly free. In the low energy effective theory the induced vacuum expectation value (vev) of the same scalar 
triplet gives rise to sub-eV Majorana masses, as required by oscillations experiments, to the three active 
neutrinos through the lepton number violating interaction $\Delta_i L L + \Delta_i^\dagger H H$, where $H$ 
is the SM Higgs. 

\section{The Scalar Triplet Model with Inelastic Dark Matter}\label{sec:model}

In addition to the vector-like lepton doublet $\psi^T \equiv (\psi^0, \psi^-)$, we have two scalar triplets $\Delta_{1,2}$. 
In our convention the scalar triplet is defined as $\Delta = (\Delta^{++}, \Delta^{+},\Delta^0)$, with hypercharge $Y=2$. 
Since the hypercharge of $\Delta$ is 2, it can have bilinear couplings to Higgs doublet $H$ as well as to the lepton doublets $L$ 
and $\psi$. The scalar potential involving $\Delta$ (from here onwards we drop the subscripts for the two scalar triplets and refer 
to them loosely as $\Delta$) and $H$ can be written as follows:
\begin{eqnarray}\label{eq:ScalarPotential}
\hspace*{-0.5cm}
 V(\Delta, H) & = &   M_\Delta^2 \Delta^\dagger \Delta + \frac{\lambda_\Delta}{2} (\Delta^\dagger \Delta)^2 -  M_H^2 H^\dagger H 
+  \frac{\lambda_H}{2} (H^\dagger H)^2 + \lambda_{\Delta H} H^{\dagger} H \Delta^\dagger \Delta \nonumber\\
&+& \frac{1}{\sqrt{2}} \left[ \mu_H \Delta^\dagger H H + {\rm h.c.}\right]  \,.
\end{eqnarray}
Below electroweak phase transition the scalar triplet acquires an induced vacuum expectation value (vev):
\begin{equation}
\langle \Delta \rangle = -f_H \frac{v^2}{\sqrt{2} M_\Delta}\,,
\end{equation}
where $v=\langle H \rangle$ = 246 GeV. The value of $\langle \Delta \rangle$ is upper bounded to be around 1 GeV in 
order not to spoil the SM prediction: $\rho \approx 1$. The bi-linear couplings of leptons and Higgs to $\Delta$ are given by:
\begin{equation}\label{eq:Lag-DM}
\hspace*{-0.5cm}
 -\mathcal{L}  \supset  \frac{1}{\sqrt{2}} \left[ f_H M_\Delta \Delta^\dagger H H + (f_L)_{\alpha,\beta} \Delta 
L_\alpha L_\beta  + f_\psi \Delta \psi \psi + {\rm h.c.} \right]\,,
\label{triplet-decay}
\end{equation}
where $f_H=\mu_H/M_\Delta$ and $\alpha, \beta=1,2,3$. The above Lagrangian satisfy a discrete symmetry $Z_2$ under which 
$\psi$ is odd, while rest of the fields are even. As a result the neutral component of $\psi$, i.e. $\psi^0 \equiv 
\psi_{\rm DM}$  behaves as a candidate of DM. When $\Delta$ acquires a vev, the $\Delta L_\alpha  L_\beta$ coupling gives 
Majorana masses to three flavors of active neutrinos as:
\begin{equation}
(M_\nu)_{\alpha \beta}= \sqrt{2} f_{\alpha\beta}\langle \Delta \rangle = -f_{L,\alpha\beta}f_H \frac{v^2}{ M_\Delta}\,.
\end{equation}
Taking $M_\Delta \sim 10^{10}$ GeV and $f_H\sim 1$ and $f_L \sim {\mathcal O}(10^{-4})$ we get $M_\nu \sim {\mathcal O}$(eV), which is 
compatible with the observed neutrino oscillation data. Moreover, the $\Delta \psi \psi$ coupling also give a Majorana mass to $\psi_{\rm DM}$ given by 
\begin{equation}
m= \sqrt{2} f_\psi \langle \Delta \rangle = f_\psi f_H \frac{v^2}{ M_\Delta}\,.
\end{equation}
Therefore the Dirac spinor $\psi_{\rm DM}$ can be written as sum of two Majorana spinors $(\psi_{\rm DM})_L$ and 
$(\psi_{\rm DM})_R$. The Lagrangian for the DM mass becomes: 
\bea
-\mathcal{L}_{\rm DM mass} &=& M_D \left[ \overline{(\psi_{\rm DM})_L} (\psi_{\rm DM})_R 
+ \overline{ (\psi_{\rm DM})_R} (\psi_{\rm DM})_L \right] \nonumber\\
&& + m \left[ \overline{ (\psi_{\rm DM})_L^c} (\psi_{\rm DM})_L + \overline{ (\psi_{\rm DM})_R^c} (\psi_{\rm DM})_R \right] \,.
\eea
This implies there is a $2\times 2$ mass matrix for the DM in the basis $\{(\psi_{\rm DM})_L, (\psi_{\rm DM})_R\}$. By diagonalising 
it two mass eigenstates $(\psi_{\rm DM})_1$ and $(\psi_{\rm DM})_2$ arise, with masses $M_{\psi_1}=M_D -m$ and $M_{\psi_2}=M_D + m$. 
The small mass splitting between the two mass eigen states:$\delta = 2 m$  led to the property of DM to be inelastic type. From the 
direct search experiments this is required to be ${\cal O} (100)$ keV. We will come back to this issue while discussing inelastic 
scattering of DM with nucleons.  

\section{Simultaneous generation of visible and DM asymmetries}
\begin{figure}
\epsfig{figure=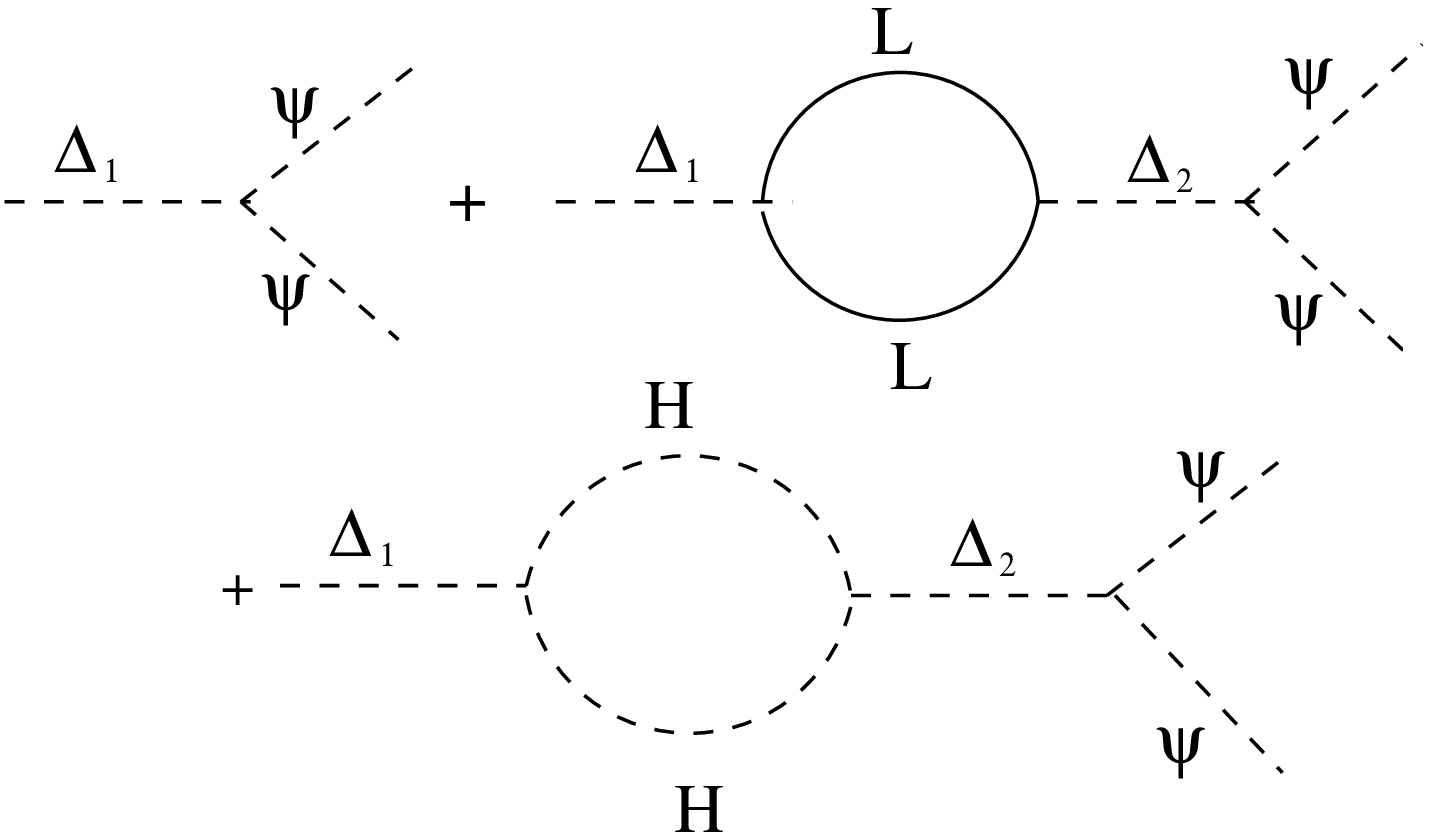,height=1.5in}
\epsfig{figure=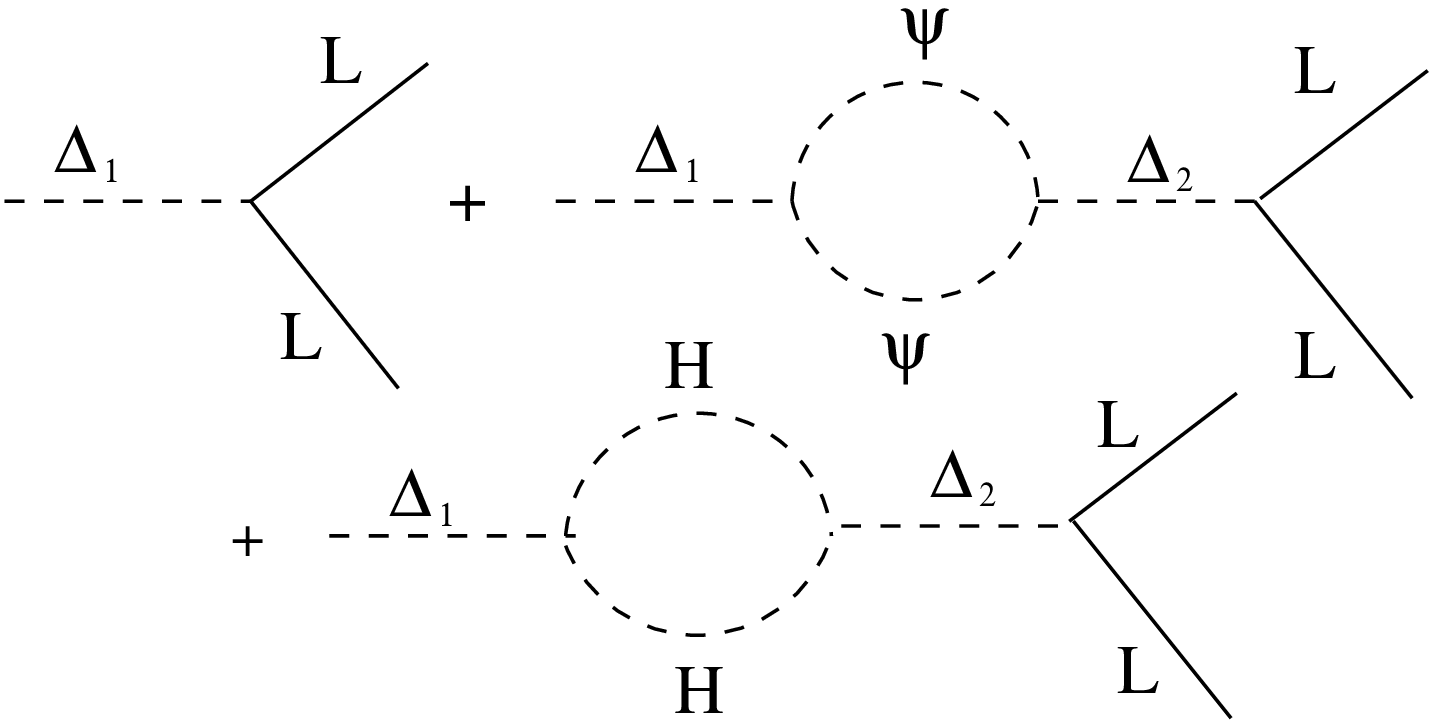,height=1.5in}
\caption{CP-violation in the dark matter and visible sectors arise through the interference of tree level with one loop self energy 
correction diagrams.
\label{fig:cpviolation}}
\end{figure}
We assume that the scalar triplets $\Delta_1$ and $\Delta_2$ are super heavy. So that the CP-violating out-of-equilibrium 
decay of the lightest one, say $\Delta_1$, in the early Universe can generate asymmetries simultaneously in visible and DM 
sectors. From Eq. (\ref{triplet-decay}) we see that the scalar triplets can decay in three channels: $\Delta\to HH$, 
$\Delta\to LL$ and $\Delta \to \psi\psi$. Moreover, these couplings are complex. Therefore, CP violation can arise from the 
interference of tree level and self-energy correction diagrams as shown in the figure (\ref{fig:cpviolation}).  
From these diagrams we see that to generate a net CP asymmetry at least two scalar triplets $\Delta_1$ and $\Delta_2$ 
are required. As a result the interaction of $\Delta_1$ and $\Delta_2$ is described by a complex mass matrix instead of 
a single mass term as mentioned in (\ref{eq:ScalarPotential}). The diagonalisation of the flavour basis spanned by 
($\Delta_1, \Delta_2$) gives rise to two mass eigenstates $\zeta^+_{1,2} = A^+_{1,2} \Delta_1 + B^+_{1,2} \Delta_2$ with 
masses $M_1$ and $M_2$. The complex conjugate of $\zeta^+_{1,2}$ are given by $\zeta^-_{1,2} = A^-_{1,2} \Delta_1 + B^-_{1,2} \Delta_2$. 
Unlike the flavor eigenstates $\Delta_1$ and $\Delta_2$, the mass eigenstates $\zeta^+_{1,2}$ and $\zeta^-_{1,2}$ are not CP 
eigenstates and hence their decay can give rise to CP asymmetry. Assuming a mass hierarchy in the mass eigenstates of the triplets, the 
final asymmetry arises by the decay of lightest triplet $\zeta_1^+$ and $\zeta_1^-$. The CP asymmetries are estimated to be:
\bea
\epsilon_{L} = & \, \, \frac{1}{8\pi^2} \frac{M_1 M_2}{M_2^2-M_1^2} \left[\frac{M_1}{\Gamma_{1}} \right]
  {\rm Im}\left[  \left( f_{1\psi} f_{2\psi}^* + f_{1H} f_{2H}^*\right) \sum_{\alpha \beta} 
(f_{1L})_{\alpha\beta} (f^*_{2L})_{\alpha\beta} \right]\,,
\label{cp_vis}
\\
\epsilon_{\rm DM} = &\, \,  \frac{1}{8\pi^2} \frac{M_1 M_2}{M_2^2-M_1^2} \left[\frac{M_1}{\Gamma_{1}} \right] 
{\rm Im} \left[ f_{1\psi} f_{2\psi}^* 
\left( f_{1H} f_{2H}^* + \sum_{\alpha \beta} (f_{1L})_{\alpha\beta} (f^*_{2L})_{\alpha\beta}
\right) \right] \,,
\label{cp_dark}
\eea
where 
\begin{equation}
\Gamma_1=\frac{M_1}{8\pi} \left( |f_{1H}|^2 + |f_{1\psi}|^2 + |f_{1L}|^2 \right)\,,
\end{equation}
is the total decay rate of the lightest triplet. When $\Gamma_1$ fails to compete with the Hubble expansion scale of the 
Universe, $\zeta_1$ decays away and produces asymmetries in either sectors. As a result the yield factors are given by:
\bea
Y_L  \equiv &\, \,  \frac{n_L}{s} = \epsilon_L X_\zeta \eta_L\,, 
\\
Y_{\rm DM} \equiv &\, \,  \frac{n_\psi}{s}  =  \epsilon_{\rm DM} X_\zeta \eta_{\rm DM}\,,
\label{asymmetry_L_DM}
\eea
where $X_\zeta = n_{\zeta_1^-}/s \equiv n_{\zeta_1^+}/s$, $s=2\pi^2 g_* T^3/45$ is the entropy density and $\eta_{L}$, $\eta_{\rm DM}$ 
are the efficiency factors, which take into account the depletion of asymmetries due to the number violating processes involving $\psi$, 
$L$ and $H$. At a temperature above EW phase transition a part of the lepton asymmetry gets converted to the baryon asymmetry via the $SU(2)_L$ 
sphaleron processes. As a result the baryon asymmetry is $Y_B = -0.55 Y_L$. From (\ref{asymmetry_L_DM}) we get the DM to baryon ratio:
\begin{equation}\label{eq:IMP}
\frac{\Omega_{\rm DM}}{\Omega_B} = \frac{1}{0.55 }\frac{m_{\rm DM}}{m_p} \frac{\epsilon_{\rm DM}}{\epsilon_L}
\frac{\eta_{\rm DM}}{\eta_L}\,,
\end{equation}
where $m_p\sim 1$ GeV is the proton mass. From this equation it is clear that the criteria $\Omega_{\rm DM} \sim 5\  \Omega_B$ can be satisfied by 
adjusting the ratio of CP-violation: $\epsilon_{\rm DM}/\epsilon_L$ and the ratio of efficiency factors: $\eta_{\rm DM}/\eta_L$. The details of 
numerical analysis can be found in refs.\cite{Arina:2011cu,Arina:2012fb}.

\section{Inelastic Inert Fermion Doublet DM and Direct Searches}

We now comment on the implications of our model for DM search. As already mentioned, the coupling between $\psi$ and $\Delta$ provides a 
small Majorana mass to $\psi_{\rm DM}$. In the mass basis, $(\psi_{\rm DM})_1$ has an off diagonal coupling with the $Z$ boson,  preventing 
it to be excluded by direct detection searches. If the mass splitting is of the order of several keV, the DM $(\psi_{\rm DM})_1$ actually has 
enough energy to scatter off nuclei and to go into its excited state $(\psi_{\rm DM})_2$, which is the definition of inelastic scattering~\cite{TuckerSmith:2001hy}. 
\begin{figure}[htbp]
\epsfig{figure=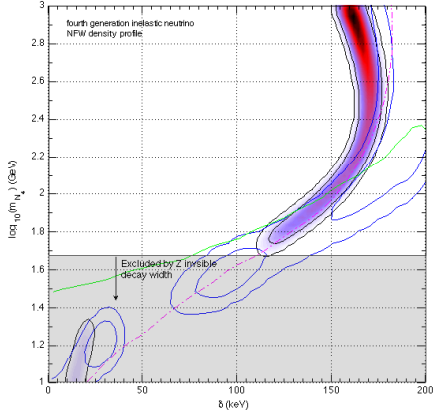, width=0.5\textwidth}
\caption{2D marginal posterior pdf in the $\{ \delta,m_{\rm DM} \}$-plane. The shaded (blue solid) contours denote the 90\% and 99\% credible regions for DAMA (CRESST) respectively. The magenta dot-dashed line is the XENON100 exclusion limit, while the green dashed line is the upper bound of KIMS experiment, at $90_S\%$ confidence level. All the astrophysical uncertainties and nuisance parameters have been marginalized over. The light gray region is excluded by LEP.
\label{fig:direct_search}}
\end{figure}
The state of art for inelastic inert fermion doublet DM is given in figure~\ref{fig:direct_search} in the $\{\delta, m_{\rm DM}\}$-plane, where the cross-section is 
fixed by the model, while the Majorana mass is allowed to vary in a reasonable range of values, in order for the scattering to occur. A Majorana mass of the 
order of 100 keV accounts for the DAMA~\cite{Bernabei:2010mq} annual modulated signal (shaded region), while a much wider range accounts for the event excess 
seen in CRESST~\cite{Angloher:2011uu} (blue non filled region). However those regions are severely constrained by XENON100~\cite{xenon:2012nq} 
and KIMS~\cite{Kim:2012rz}. KIMS is very constraining being a scintillator with Iodine crystals as DAMA. Our DM candidate can explain simultaneously 
the DAMA and CRESST detection, with a  marginal compatibility at $90_S\%$ with XENON100 and KIMS, for a mass range that goes from 45 GeV up to $\sim 250$ GeV. 
If we give up the DAMA explanation, then it could account for the CRESST excess up to masses of the order of $\sim 500$ GeV. 

In summary we present a scalar triplet extension of the SM where the decay of triplet in the early Universe produces visible and DM, while its induced 
vev give rise to Majorana masses to three flavours of active neutrinos in the late universe.  

\section*{Acknowledgments}
I would like to thank Rabindra N. Mohapatra and Chira Arina for useful discussions.

\end{document}